\documentclass{emulateapj}
\usepackage{amsmath}
\usepackage{lineno}

\shorttitle{Hard-sphere-like Acceleration in Blazars}
\shortauthors{Asano \& Hayashida}

\begin{document}

\title{
Blazar Spectra with Hard-sphere-like Acceleration of Electrons
}

\author{Katsuaki Asano\altaffilmark{1}, and Masaaki Hayashida\altaffilmark{2}}

\affil{\altaffilmark{1}Institute for Cosmic Ray Research, The University of Tokyo, 5-1-5 Kashiwanoha, Kashiwa, Chiba 277-8582, Japan; asanok@icrr.u-tokyo.ac.jp}
\affil{\altaffilmark{2}Department of Physics, Faculty of Science and Engineering, Konan University, 8-9-1 Okamoto, Kobe, Hyogo 658-8501, Japan; mahaya@center.konan-u.ac.jp}


\begin{abstract}
Electrons emitting non-thermal photons in blazars are possibly accelerated
by turbulences developed in jets.
In this paper, we consider the case so-called hard-sphere scattering
as an interaction model between turbulences and electrons, in which
the acceleration timescale is independent of the electron energy.
We numerically simulate broad-band emission from blazar jets
with a one-zone time-dependent code, taking into account the
turbulence acceleration.
Our model reproduces various blazar spectra with simple assumptions,
such as constant particle injection rate, constant diffusion coefficient,
and conical geometry of the jet.
We also discuss possible mechanism to realize the hard-sphere-like acceleration
in blazar jets.
\end{abstract}


\section{Introduction}
\label{sec:intro}

The double-peak structure of blazar spectra is well modeled with
leptonic models
\citep[e.g.][]{ghi85,der93,sik94,tav98,bot00,kin02,cel08,bot13}.
In the leptonic models, gamma-ray emission is
generated via inverse Compton (IC) scattering.
The synchrotron-self Compton (SSC) process up-scatters
synchrotron photons produced by the same electron population
in jets \citep[e.g.][]{mar92}.
Since Flat Spectrum Radio Quasars (FSRQs) have more luminous disk, external photons surrounding
the disk dominate the seed photons for the IC scattering \citep[e.g.][]{sik94}.
In most models, the non-thermal electrons are assumed to be accelerated
via shocks with the Fermi process.
However, there are several unsolved problems in the particle acceleration
mechanism in blazars.
Some blazar spectra imply that the electron power-law index is smaller
than 2, the value from the simplest shock acceleration model.
The change of the index at the spectral break is
very large inconsistently with the cooling break \citep[see e.g.][]{hay12,yan14}.
To reproduce the curved photon spectra, an emission model requires an
electron energy distribution with many phenomenological parameters,
such as a double-broken power-law with a low-energy cutoff
\citep[e.g.][]{abd11}.

The maximum electron energy in blazars is much smaller than
the value implied from the Bohm limit \citep{ino96}.
The acceleration timescale in blazars
seems to be very long compared to the timescale of the shock acceleration
considered in supernova remnants \citep{aha99} or pulsar wind nebulae \citep{ato96}.

Also from the theoretical point of view, the shock acceleration in blazars
has difficulties.
To dissipate significant kinetic energy consistently with
blazar luminosity, a relativistic shock is favored.
However, a relativistic shock tends to become superluminal
\citep{kir89}, where electrons moving along the magnetic field
in the downstream cannot return to the upstream
so that the shock acceleration is inhibited \citep{sir15}.
Furthermore, the shock acceleration can operate only in weakly
magnetized plasma \citep{lem11}, which may contradict
the magnetization implied from the spectral modeling \citep[e.g.][]{tav98,hay15}.
This is because, for a relativistic shocks propagating in a marginally magnetized plasma,
the turbulence in the upstream as the shock precursor is hard to be excited by the particles
reflected from the shock front.
The absent of the precursor turbulence in the upstream prohibits electrons
from returning to the downstream.

As for the spectral hardness in the shock acceleration theory,
the nonlinear effect of the cosmic-ray pressure may modify the shock structure
and harden the electron spectrum \citep{mal01}.
However, a Monte Carlo simulation by \citet{ell13}
shows that the nonlinear effect of the cosmic-ray
on the shock structure weakens, and the acceleration efficiency
is also suppressed, as the shock speed
becomes relativistic.

An alternative candidate of acceleration mechanism
is the stochastic acceleration by turbulence \citep[e.g.][]{sch84,par95,bec06,cho06,sta08},
which is phenomenologically equivalent to the second-order Fermi acceleration.
The required turbulence may be excited by Kelvin--Helmholtz instability \citep{ros08}
at the boundary of the spine--sheath structure \citep{ghi05},
Rayleigh--Taylor and Richtmyer--Meshkov instabilities in the transverse jet structure
\citep{mat13}, kink instability \citep{bro16}, tearing mode instability \citep{sir14},
or the star--jet interaction \citep{bar12}.

The turbulence acceleration is slower process than the shock acceleration.
In addition, this process can produce a hard electron spectrum
with an index smaller than 2.
Given the energy diffusion coefficient $D_{\varepsilon \varepsilon} \propto \varepsilon^q$,
the number spectrum in the steady state without the escape effect
is proportional to $\varepsilon^{1-q}$, where $\varepsilon$ is the particle energy.
If the particle scattering is dominated by the gyro-resonant scattering with
turbulent waves,
the power-law index $q$ in $D_{\varepsilon \varepsilon}$ is determined
by the magnetic wave spectrum as $\delta B^2(k) \propto k^{-q}$
\citep{bla87}.

The turbulent acceleration model has been adopted to blazars
by many authors \citep[e.g.][]{bot99,sch00,kak15,asa14,asa15}.
However, \citet{kak15} shows that a very extreme parameter set,
such as tiny emission region,
is required to reproduce the spectra of Mrk 421 and Mrk 501
by the steady state one-zone model.
The results in \citet{kak15} imply that the steady
state of the electron energy distribution is hard to be realized
by the balance among the turbulence acceleration, injection,
cooling, and escape of electrons.
In the slow acceleration process,
the temporal evolution of the electron energy density along the jet
is essential to discuss the resultant photon spectrum.
Adopting the Kolmogorov index ($q=5/3$), \citet{asa14} shows
that a model taking into account the radial evolution of the
electron energy distribution reconciles the observed spectrum of Mrk 421,
but the model in \citet{asa14} requires a non-trivial evolution
of the electron injection rate.

On the other hand, the model in \citet{asa15} succeeds in reproducing
the spectra for both the steady and flare states of 3C 279
with a simple assumption and the hard-sphere index $q=2$,
with which the acceleration
timescale is independent of the particle energy.

In this paper, we pursue the hard-sphere-like acceleration in blazars
by fitting spectra of five representative blazars, Mrk 421, 1ES 1959+650,
PKS 2155-304, 3C 279 and PKS 1510-080, taking into account the radial evolution
of the electron energy distribution.
The sample consists of both types of blazars, BL Lac objects
and FSRQs. 
Blazars emission often shows strong variability.
Thanks to recent improvements of observational instruments,
broad-band data from contemporaneous observations become available
not only in high-flux flaring states but also in general steady states of the blazars.
We demonstrate that the broad-band spectra of those blazars
in the steady states are reproduced with the hard-sphere-like acceleration model
with a small number of parameters.

In \S \ref{sec:meth}, we explain our model and method to
produce model spectra.
The results for several blazars are summarized in \S \ref{sec:res}.
\S \ref{sec:dis} is devoted to discuss the implication
for the acceleration mechanism in blazars.

\section{Method}
\label{sec:meth}

In this paper, the steady state photon spectrum is calculated
by the numerical code in \citet{asa14}
\citep[see also,][]{asa15}.
To explain the model parameters summarized in Table \ref{param_tab},
we shortly review the calculation method below.
The steady emission from a blazar is modeled with a steady
outflowing jet, for which we consider continuous shell ejection
from the initial radius $R_0$.
Although a conical jet with an opening angle
of $1/\Gamma$ is assumed to calculate the emission,
where $\Gamma$ is the bulk Lorentz factor of the jet,
the electron injection rate $\dot{N}'$ is normalized
by a spherically-equivalent volume,
\begin{equation}
V'_0=4 \pi R_0^3/\Gamma,
\end{equation}
at $R=R_0$. Hereafter, we denote values in the comoving frame
by prime characters.
The volume expands as $V' \propto R^2$ for the conical geometry.
The electron injection rate into the above reference volume $\dot{N}'$
is assumed as constant during the expansion timescale.
Then, the injection rate density behaves as
$\dot{n}'=\dot{N}'/V' \propto R^{-2}$ from $R=R_0$ to $2 R_0$.
The electron injection is monoenergetic with an initial
Lorentz factor of $\gamma'_{\rm inj}$.
We have confirmed that a power-law injection
yields a harder electron spectrum than the examples
in this paper. Since our purpose is to yield a soft spectrum
with the turbulence acceleration model,
we assume the monoenergetic injection.

The particle acceleration by turbulences
is regulated by the energy diffusion coefficient,
\begin{equation}
D'_{\varepsilon \varepsilon}=K \varepsilon'^2_{\rm e},
\label{difdif}
\end{equation}
where $\varepsilon'_{\rm e}$ is the electron energy.
In this paper, we consider only the hard-sphere type
acceleration as defined in equation (\ref{difdif}).
The parameter $K$ is constant for $R \leq 2 R_0$.

The radial evolution of the electron energy distribution
is calculated taking into account the electron injection,
turbulence acceleration, radiative cooling, and adiabatic cooling.
For $R > 2 R_0$ the injection and energy diffusion shut down.
Our numerical code follows the evolution of the electron energy distribution
and photon production as far as $R=R_{\rm out}$.
The synchrotron emissivity and self-absorption are calculated
with the magnetic field evolving as
\begin{eqnarray}
B'=B_0 \left( \frac{R}{R_0} \right)^{-1},
\end{eqnarray}
where $B_0$ is the initial value.

In addition to SSC,
IC emission due to the external photon field (EIC)
is calculated for FSRQs.
The models for the external photon fields are the same as discussed in \citet{hay12}
with the broad line emission and the infrared dust emission.
When the broad line region (BLR) is considered as the external photon source,
the energy density is written as
\begin{eqnarray}
U'_{\rm UV}=\frac{0.1 \Gamma^2 L_{\rm D}}{3 \pi c R^2_{\rm BLR} (1+(R/R_{\rm BLR})^3)},
\end{eqnarray}
where
$L_{\rm D}$ is the disk luminosity, and
the size of BLR is described as
\begin{eqnarray}
R_{\rm BLR}=0.1 \left( \frac{L_{\rm D}}{10^{46}~\mbox{erg}~\mbox{s}^{-1}} \right)^{1/2}~\mbox{pc}.
\end{eqnarray}
The external photon spectrum from BLR is approximated by the diluted Planck distribution
with photon temperature of
\begin{eqnarray}
T'_{\rm UV}=10 \Gamma~\mbox{eV}.
\end{eqnarray}

The infrared dust emission is an alternative source of the external photon field
for a relatively larger $R_0$.
In this case, we adopt a steeply dropping function as
\begin{eqnarray}
U'_{\rm IR}=\frac{0.1 \Gamma^2 L_{\rm D}}{3 \pi c R^2_{\rm IR} (1+(R/R_{\rm IR})^4)}.
\end{eqnarray}
The parameters in this case are expressed as
\begin{eqnarray}
R_{\rm IR}&=&2.5 \left( \frac{L_{\rm D}}{10^{46}~\mbox{erg}~\mbox{s}^{-1}} \right)^{1/2}~\mbox{pc},\\
T'_{\rm IR}&=&0.3 \Gamma~\mbox{eV}.
\end{eqnarray}

In summary, the model parameters are $\Gamma$, $R_0$, $B_0$, $\gamma'_{\rm inj}$, $K$,
$R_{\rm out}$, and $L_{\rm D}$.
The shape of the electron spectrum is adjusted by only the diffusion coefficient
$K$ and the cooling effect.
The conically expanding outflow in this model seems reasonable and simplest
assumption. We do not need to put an additional parameter for particle escape.
The adiabatic cooling effect suppresses the emission at $R \gg R_0$,
so that the steady outflow results in a steady emission without the escape effect.

\section{Results}
\label{sec:res}

\begin{table*}[!hbtp]
	\caption{Model parameters}
	\begin{center}
		\begin{tabular}{lcccccccccc}
			\hline\hline
			 & & $\gamma'_{\rm inj}$ & $R_0$ & $R_{\rm out}/R_0$ & $\Gamma$ & $B_0$  & $K$ & $\dot{N}'$ & $L_{\rm D}$ & UV/IR \\
			 & & & cm & & & G & $\mbox{s}^{-1}$ & $\mbox{s}^{-1}$ & erg $\mbox{s}^{-1}$ & \\
			\hline
			Mrk 421 & A &  10 & $1.5 \times 10^{17}$ & 30 & 15 & 0.18 & $4.8 \times 10^{-6}$ & $2.4 \times 10^{47}$ & --- & --- \\
			 & B &  100 & $1.5 \times 10^{17}$ & 30 & 15 & 0.16 & $3.7 \times 10^{-6}$ & $9.8 \times 10^{46}$ & --- & --- \\
            \hline
			1ES 1959+650 & A & 100 & $1.6 \times 10^{17}$ & 30 & 20 & 0.18 & $5.0 \times 10^{-6}$ & $3.7 \times 10^{46}$ & --- & --- \\
			 & B & 100 & $1.6 \times 10^{17}$ & 10 & 20 & 0.18 & $5.0 \times 10^{-6}$ & $3.7 \times 10^{46}$ & --- & --- \\
			 & C & 10 & $4.0 \times 10^{16}$ & 30 & 40 & 0.5 & $4.3 \times 10^{-5}$ & $1.5 \times 10^{47}$ & --- & --- \\
			\hline
			PKS 2155--304 &   & 10 & $6.0 \times 10^{16}$ & 30 & 20 & 1.2 & $1.2 \times 10^{-5}$ & $1.5 \times 10^{48}$ & --- & --- \\
			\hline
			3C 279 &   & 10 & $7.1 \times 10^{16}$ & 30 & 15 & 8.0 & $9.5 \times 10^{-6}$ & $7.3 \times 10^{49}$ & $6.0 \times 10^{45}$ & UV \\
			\hline
			PKS 1510--089 &   & 10 & $6.0 \times 10^{17}$ & 30 & 20 & 0.38 & $9.0 \times 10^{-7}$ & $7.3 \times 10^{49}$ & $5.0 \times 10^{45}$ & IR \\
			\hline\hline
		\end{tabular}
	\end{center}
	\label{param_tab}
\end{table*}

We select several spectral data for the blazars in non-flaring states.
The model parameters we obtained
are summarized in Table \ref{param_tab}.
In all the cases, the acceleration timescale $K^{-1}$
is comparable to the dynamical timescale $R_0/(\Gamma c)$.
A slight difference in $K$ results in a drastic change
of the resultant photon spectrum.
Below we discuss the individual cases.

\subsection{Mrk 421}
\label{sec:Mrk421}

First, we discuss a famous BL Lac object Mrk 421
(redshift $z=0.031$)
based on the spectrum of the 4.5 month long multifrequency
campaign \citep{abd11}.
For this data set, \citet{asa14} and \citet{kak15} have already
fitted the spectrum with turbulence acceleration models.
The steady one-zone model of \citet{kak15} requires
a nontrivial energy-dependence of the diffusion coefficient,
$D'_{\varepsilon \varepsilon} \propto \varepsilon'^{1.85}_{\rm e}$,
and very small region ($\sim 10^{14}$ cm) to make electrons escape
faster.
On the other hand, the model in \citet{asa14} is the same as
the model in this paper, but the diffusion coefficient
was assumed as the Kolmogorov type,
$D'_{\varepsilon \varepsilon} \propto \varepsilon'^{5/3}_{\rm e}$.
To reproduce the broad photon spectrum,
a rapid evolution of the particle injection with radius
was required.
In this section, we demonstrate that the hard-sphere model
can reproduce the spectrum even with a constant injection rate.
The macroscopic parameters $R_0$ and $\Gamma$ in our model
roughly agree with the values in the original work on the campaign data \citep{abd11}.
The variability timescale $\sim R_0/\Gamma^2$ corresponds
to about $0.3$ day.

\begin{figure}[!h]
\centering
\epsscale{1.0}
\plotone{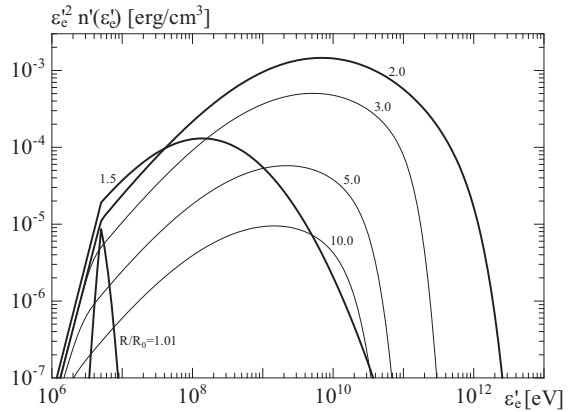}
\caption{Evolutions of the energy distribution of the electron density
in the model A for Mrk 421.}
\label{Ele}
\end{figure}

Figure \ref{Ele} shows the radial evolution of the electron energy
distribution for model A (see Table \ref{param_tab}).
Here, we inject electrons with the initial Lorentz factor $\gamma'_{\rm inj}=10$.
In the hard-sphere case, the electron energy distribution
is sensitive to the ratio of the acceleration timescale $K^{-1}$
to the elapsed time $t'$ as analytically demonstrated in \citet{asa16}.
The energy injected from turbulence grows exponentially
as $\propto \exp{(Kt')}$ for $Kt' \geq 1$.
In the low-energy part above $\gamma'_{\rm inj} m_{\rm e} c^2$,
the spectrum can be approximated as a power-law of $n'(\varepsilon'_{\rm e})
\propto \varepsilon'^{-1}_{\rm e}$.
The spectral shape around the peak is curved, which is characteristic
for the stochastic acceleration.
After the shutdown of the electron acceleration and injection
($R>2R_0$), we can see that the adiabatic cooling effect shifts
the spectral shape to lower energy.

\begin{figure}[!h]
\centering
\epsscale{1.0}
\plotone{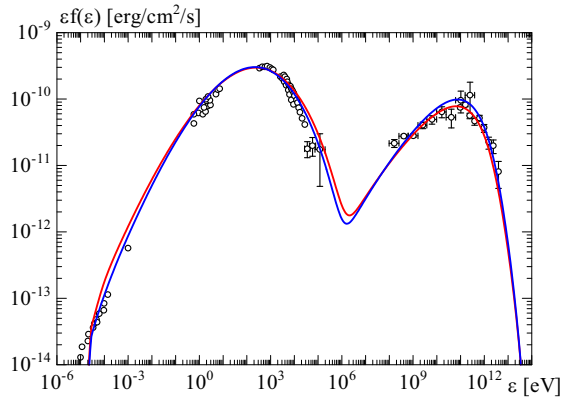}
\caption{Model photon spectra A (red) and B (blue) for Mrk 421.
The observed data points are partially extracted from the data of the 4.5 month
campaign \citep{abd11}.}
\label{Mrk421}
\end{figure}

Owing to the curved electron spectrum and its radial evolution,
the curved spectrum for Mrk 421 is well reproduced even from radio
to gamma-ray as shown by the red line in Figure \ref{Mrk421}.
The synchrotron spectrum is broad enough to agree with the observation
differently from the Kolmogorov case in \citet{asa14}.
Although the radio data seems on the extrapolation from the optical/IR data,
the radio emission has been neglected as another component
in the previous models \citep[e.g.,][]{tra09,abd11},
in which absorption due to synchrotron self-absorption
is crucial in the radio band.
Owing to the hard electron spectrum in our model,
the emission region is optically thin even for radio emission.

Here, we also test model B (blue), in which
$\gamma'_{\rm inj}$ is higher.
Even for a different value of $\gamma'_{\rm inj}$,
which is highly uncertain, a slight change of the other parameters
produces a quite similar spectrum.
As shown in Figure \ref{Mrk421}, a higher $\gamma'_{\rm inj}$
tends to produce a slightly narrower spectrum.



\subsection{1ES 1959+650}
\label{sec:1ES1959}

1ES 1959+650 ($z=0.047$) is categorized as a high-frequency peaked BL
Lac object (HBL) and one of the most frequently observed TeV gamma-ray
blazars. Here we focus on a multi-wavelength campaign conducted during
a low TeV flux state in May 2006, where simultaneous X-ray data by
Suzaku and very-high-energy gamma-ray data by MAGIC are available
\citep{tag08}.

\begin{figure}[!h]
\centering
\epsscale{1.0}
\plotone{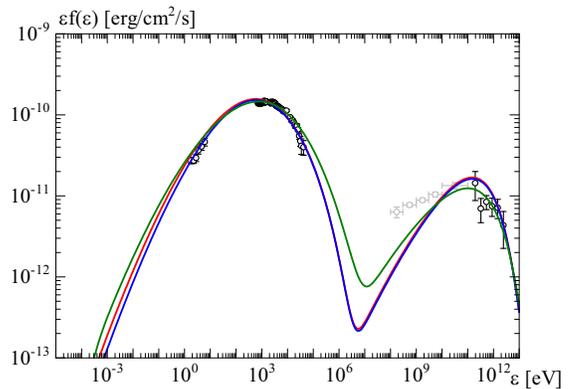}
\caption{Model photon spectra A (red), B (blue) and C (green) for 1ES 1959+650.
The black open circles are measured flux points from the multiwavelength campaign
in 2006 May \citep{tag08}.
The gray open circles are taken from the LAT 4-year Point Source Catalog
\citep[3FGL,][]{ace15}.}
\label{1ES1959}
\end{figure}

As shown in Figure \ref{1ES1959}, the synchrotron component
shows a similar shape to that in Mrk 421.
Compared to the spectrum of Mrk421,
the spectral peak energies of the synchrotron and SSC components are similar
while the ratio of the gamma-ray flux to the synchrotron flux in 1ES
1959+650 is relatively small.
It suggests that the typical electron energy and magnetic field may
be similar in the two objects,
but the efficiency of the SSC emission in 1ES1959+650 needs to be
reduced by increasing the bulk Lorentz factor slightly.

Our results are summarized in Figure \ref{1ES1959},
in which the parameter values of $R_0$, $\Gamma$, and $B_0$
are basically consistent with the previous results in \citet{tag08}.
Model A (red) approximately reproduces the curved
spectral shape for the synchrotron component,
but the IR/optical flux is slightly higher than the observed flux.
This difference may be within theoretical uncertainty in our idealized model.
For instance, since the magnetic field may be not purely toroidal,
the field may decay faster than $B' \propto R^{-1}$.
In such a case, the synchrotron emission for a large $R$ would
be suppressed.
As model B (blue, where $R_{\rm out}=10 R_0$) shows,
if the emission from $R>10 R_0$ is neglected,
the synchrotron spectrum becomes more consistent with the observed data.
The slight difference in the spectra below 10 eV between models A and B
exhibits the small contribution of emission at $R>10 R_0$.

While the spectral shape of the synchrotron component
in model A or B seems consistent with the observation,
the gamma-ray spectrum may contradict the observed date
in the GeV energy range.
Note that the gray data points in the GeV range shown in Figure \ref{1ES1959} are
4-year averaged spectral data,
and they were not simultaneously obtained with the data in other energy ranges.
The variability in the GeV energy range \citep[e.g.,][]{kau17,pat18}
may resolve the discrepancy.
If we seriously incorporate the soft spectrum indicated by
the {\it Fermi} data points,
a broader energy distribution is required for electrons.
Model C (green), in which a smaller $\gamma'_{\rm inj}$ and higher $\Gamma$
are adopted, shows a broader gamma-ray spectrum.
Even in this case, the model flux below 1 GeV
is significantly lower than the observed data,
and the synchrotron spectrum becomes
broader than the data.
To reconcile both the narrow synchrotron and soft IC component,
an external photon field may be required as discussed
in \citet{asa14}.

\subsection{PKS 2155--304}
\label{sec:PKS2155}

PKS 2155--304 ($z=0.116$) is one of the most luminous HBL
objects in the TeV energy band.
The object is famous for short-timescale variability ($\sim$ minutes)
in the TeV gamma-rays. Recently, a spectral hardening in the hard
X-ray band has been discovered from observations with the NuSTAR satellite
\citep{mad16}. The spectral hardening can be interpreted by
the onset of the IC component. The observations were performed in a
very low X-ray state together with XMM-{\it Newton} and {\it Swift}-UVOT, which provide
soft X-ray and UV data for the synchrotron component.
We apply our model for the broad band spectrum
also including gamma-ray data by {\it Fermi}-LAT as reported in \citet{mad16}.

\begin{figure}[!h]
\centering
\epsscale{1.0}
\plotone{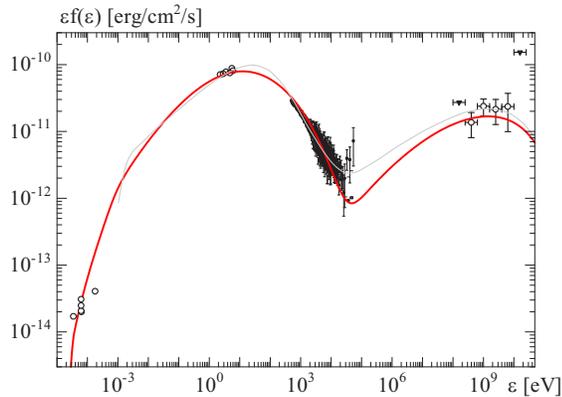}
\caption{Model photon spectrum (red) for PKS 2155--304.
The open circles for optical, X-ray and gamma-ray are fluxes measured in 2013
April \citep{mad16}.
The gray thin line is the broken power-law model in \citet{mad16}.
The radio data points represent core fluxes taken from NASA/IPAC Extragalactic Database (NED)}
\label{PKS2155}
\end{figure}

As the thin grey line in Figure \ref{PKS2155} shows,
the photon spectrum is fitted by a model with
a broken power-law electron distribution (the low-energy index
is 2.2, high-energy index is 3.8, and the break Lorentz factor
is $2.6 \times 10^4$) in \citet{mad16}.
However, this soft electron spectrum implies
the proton luminosity of $>10^{47}~\mbox{erg}~\mbox{s}^{-1}$
assuming the same number for protons and electrons.
\citet{mad16} concluded that the obtained proton luminosity is too large
as an HBL-type blazar.

The hard spectrum in the turbulence acceleration model
significantly suppresses the number of electrons.
The model shown in Figure \ref{PKS2155} (red) seems
consistent with the curved feature in the soft X-ray data.
In this model, the mean energy of electrons is about GeV:
the mean Lorentz factor is $\sim 2000$,
while that is $5.6$ in the model of \citet{mad16}.
Thus, we can reduce the proton luminosity by a factor of 400.

However, the model flux in the hard X-ray regime
is lower than the best-fit model of \citet{mad16},
because our model spectrum is significantly harder than
the model in \citet{mad16}.
To reconcile with the hard X-ray flux, we need another
external photon field as the IC seed photons again
\citep[see the model and discussion in][]{asa14}.

\subsection{3C 279}
\label{sec:3C279}

\begin{figure}[!h]
\centering
\epsscale{1.0}
\plotone{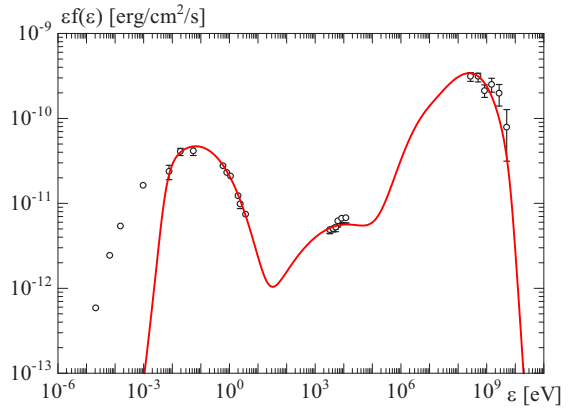}
\caption{Model photon spectrum for 3C 279.
The open circles are measured flux points in 2009 February
\citep{hay12}.}
\label{3C279}
\end{figure}

We revisit a famous FSRQ 3C 279 ($z=0.538$), while
we have tested this blazar with the turbulence acceleration model
in \citet{asa15}.
The adopted data of the spectrum
were obtained in an active period \citep[period ``D'' in][]{hay12},
but the data are averaged over five days, which may be
significantly longer than the variability timescale.
Thus, we adopt a steady emission model even in this case.
In \citet{asa15}, the external photon field was assumed to be
spatially constant.
With the photon field model explained in section \ref{sec:meth},
a similar result to \citet{asa15} is obtained as shown in Figure \ref{3C279}.

The low-energy cutoff in the radio band is due to synchrotron self-absorption.
The X-ray bump is attributed to SSC emission, while EIC emission
produces the gamma-ray peak.
The synchrotron spectral shape especially for the high-energy region
is well reproduced by our model.

\subsection{PKS 1510--089}
\label{sec:PKS1510}

To test the case of the infrared dust emission as the external photon field,
we fit the spectrum of a FSRQ PKS 1510--089 (z=0.36).
\citet{nal12} obtained broad band spectra from
multi-wavelength observations including Herschel satellite pointing
for far-infrared bands using PACS and SPIRE instruments.
They considered three components for the emission origin: jet
emissions in BLR and the hot-dust region, and
emission attributed to the accretion disk. The optical/UV bump and
X-ray components are considered to be the contributions from the
accretion disk and the hot disk corona, respectively, while
synchrotron emission is dominant in the far-infrared band observed by
Herschel.

\citet{nal12} adopted the two zone model, where the GeV emissions
are emitted from BLR and the emission from the hot-dust region
does not contribute to the gamma-ray emission so much. However,
\citet{ale14} reported detections of sub-TeV emission even during steady states and
possible correlations between GeV gamma-ray flares and radio core
appearances. Those results suggest that the gamma-ray emission is
likely to be originated outside of BLR. In this
paper, we consider the emission from only the hot-dust region;
the X-ray and optical components are not weighted so much
because of the contributions from disk and corona emission.

\begin{figure}[!h]
\centering
\epsscale{1.0}
\plotone{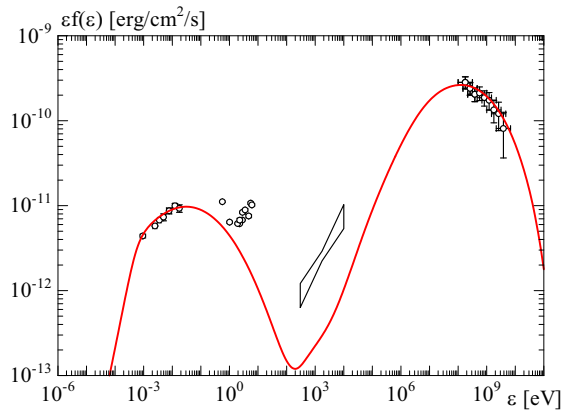}
\caption{Model photon spectrum for PKS 1510-089.
The open circles are measured flux points in 2011 August
\citep[epoch ``H2'',][]{nal12}.}
\label{PKS1510}
\end{figure}

Our model parameters in Table \ref{param_tab} imply
that the radius $R_0 \sim 0.2$ pc is
outside $R_{\rm BLR} \sim 0.07$ pc, but inside
the hot-dust region ($R_{\rm IR} \sim 1.8$ pc).
Figure \ref{PKS1510} shows our fitting results.
The gamma-ray spectral component is the EIC emission
from the hot-dust region.
The SSC emission is negligible because of the low synchrotron photon density
at the large initial radius.
The success in Figure \ref{PKS1510} suggests possible turbulence
acceleration even in the hot-dust region.

\section{Discussion}
\label{sec:dis}

The model with the hard-sphere like electron acceleration
in this paper is constructed only under simple assumptions;
constant energy diffusion coefficient,
constant injection rate of electrons, conical geometry ($V' \propto R^2$),
and $B' \propto R^{-1}$.
Nevertheless, our model spectra
have well reproduced the curved photon spectra of the blazars
from the radio to the high-energy gamma-ray bands,
though the low-energy tails of the SSC components
for 1ES 1959+650 and PKS 2155-304 would require
some modification in our model,
such as an extra photon field or a low-energy extra population of electrons.
In our model, the electron energy distribution is controlled by the three timescales:
dynamical ($R_0/\Gamma/c$), acceleration ($1/K$),
and cooling timescales.
The combinations of those parameters can produce variety
of the spectral shape without extra parameters.

\begin{figure}[!h]
\centering
\epsscale{1.0}
\plotone{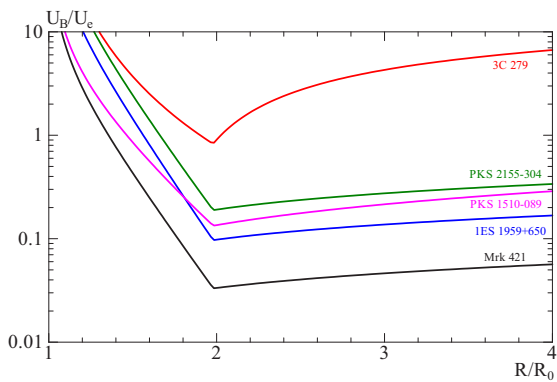}
\caption{Radial evolution of the energy density ratios for model B of Mrk 421 (black),
model B of 1ES 1959+650 (blue),
PKS 2155-304 (green), 3C 279 (red), and PKS 1510-089 (magenta).}
\label{Eneden}
\end{figure}

Here we have not specified the mechanism of the turbulence excitation
to accelerate particles.
One candidate is magnetic reconnection in highly magnetized plasma
\citep[e.g.][]{sir14,sir15,ois15,tak18}.
As \citet{sir15} demonstrated numerically,
the energy dissipation due to magnetic reconnection
leads to an equi-partition between the magnetic field and
non-thermal particles.
As shown in Figure \ref{Eneden},
the energy densities of the magnetic field and electrons
are equi-partition only for the case in 3C 279
\citep[see also,][]{der14}.
In this case, magnetic reconnection is a promising mechanism
for the turbulence excitation, or reconnection may directly
accelerate electrons, which can be equivalent to the stochastic acceleration
expressed by the diffusion coefficient in this paper.

On the other hand, the other objects in Figure \ref{Eneden},
the magnetic field is subdominant compared to the electron energy density.
Even for the case of 3C 279,
it is reported that the magnetic field energy became much lower than
the electron energy for the gamma-ray flare states \citep{hay15,asa15}.
In those cases, magnetic reconnection is unlikely
as the energy source of turbulences or non-thermal electrons.

The acceleration timescale in the hard-sphere model
does not depend on electron energy.
When particle scattering is dominated by a certain size
of eddy, the acceleration timescale will be
common irrespectively of electron energy.
However, the turbulence may be expressed by a superposition of MHD waves
in small scales.

The fast mode is the most likely dominant wave mode as the electron energy source
in low-magnetized plasma implied from our results.
The energy of the turbulence is injected at a large scale
($<R_0/\Gamma$) as fast waves, and cascade to shorter scales following
the Kolmogorov law; $k E(k) \propto k^{-2/3}$,
where $k$ is the wave number.
On the other hand, the power spectrum of the magnetic field
tends to be flatter than the kinetic one
\citep{cho00,cho09}.
At a scale, where the magnetic power is comparable
to the kinetic one,
the kinetic energy transfer mechanism in the turbulence cascade process
changes from the hydrodynamical one to
the magnetohydrodynamic one.
The power spectra for both the kinetic and magnetic energies
become steeper at this equi-partition scale \citep[$k=k_{\rm max}$,][]{gs95,ino11}.
In this case, the acceleration timescale may be regulated
by this scale.

If the Larmor radius of electrons is significantly shorter
than the wavelength $k_{\rm max}^{-1}$,
the pitch angle diffusion via gyro resonance
is not responsible for particle scattering.
In this case, the main mechanism of the energy exchange between electrons and waves
is transit time damping \citep[TTD,][]{ber58};
electrons are accelerated when their velocity along the magnetic field
equals to the parallel component of the phase velocity.
For non-relativistic waves, the number fraction of relativistic electrons that satisfy
the TTD resonance condition is small.
Only electrons with their pitch angle $\sim \pi/2$ can interact with waves.
However, the phase velocity of the wave in the blazar emission region
may be mildly relativistic.
Furthermore, the mirror force can broaden the resonance condition of TTD
as shown by \citet{yan08}.
We can expect that a significant fraction of electrons
interact with the waves.

When electrons are scattered by waves, whose wavelength are longer
than the gyro radius,
the description of the second order Fermi acceleration may be valid.
The scattering time scale is $\sim (c k_{\rm max})^{-1}$,
and the average energy change per scattering is $\Delta \varepsilon_{\rm e}/\varepsilon_{\rm e}
\sim v(k_{\rm max})/c$.
Let us assume that the turbulence velocity $v(k_{\rm max})$ at $k=k_{\rm max}$
is expressed by the Kolmogorov law as $v^2(k_{\rm max})=v_0^2 (k_{\rm max}/k_{\rm min})^{-2/3}$,
where $v_0$ is the turbulence velocity at the injection scale $k_{\rm min}^{-1}$.
Then, the acceleration frequency becomes
\begin{eqnarray}
t_{\rm acc}^{-1} \sim c k_{\rm max} \left( \frac{v(k_{\rm max})}{c} \right)^2
= c k_{\rm max} \left( \frac{v_0}{c} \right)^{2}
\left( \frac{k_{\rm max}}{k_{\rm min}} \right)^{-2/3}
\end{eqnarray}

Here we take model B for Mrk 421 as an example.
As shown in Figure \ref{Ele}, TeV is sufficient for the maximum energy of electrons
to reproduce the photon spectrum.
In this model, the Larmor radius of TeV electrons is $2 \times 10^{10}$ cm,
which should be shorter than $k_{\rm max}^{-1}$ to realize the non-gyro-resonant
scattering as dominant acceleration process.
On the other hand, we need a $k_{\rm min}^{-1}$ shorter than $R_0/\Gamma=10^{16}$ cm.
Those conditions imply $k_{\rm max}/k_{\rm min}<4.8 \times 10^5$.
With a conservative assumption of $k_{\rm min}=\Gamma/R_0$,
$c k_{\rm max} (k_{\rm max}/k_{\rm min})^{-2/3}< 2.3 \times 10^{-4}~\mbox{s}^{-1}$.
The turbulence velocity at the injection would be slower then the sound speed
in relativistic plasma: $(v_0/c)^2<1/3$.
Finally, the maximum value of $t_{\rm acc}^{-1}$ is estimated as
$7.8 \times 10^{-5}~\mbox{s}^{-1}$, which is much larger
than $K \sim 10^{-6}~\mbox{s}^{-1}$ required in the model.
If the turbulence is exicited by the star-jet interaction,
a value of $k_{\rm min}$ much larger than $\Gamma/R_0$ is possible,
which further shortens $t_{\rm acc}$.
Thus, the non-gyro-resonant scattering
may provide a mechanism to realize the hard-sphere-like
acceleration in blazars as required by our models.

Our simplest model with
the Kolmogorov-like model ($q=5/3$) cannot fit the blazar spectra.
Both the synchrotron and IC components become narrower than
the observed shape for those cases.
If the particle acceleration is due to turbulence in the jet,
the hard-sphere-like diffusion in the energy space seems necessary.
However, it is difficult to distinguish observationally
turbulence acceleration from other mechanisms.
As demonstrated in \citet{asa14}, the flare light curves may provide
a hint for the acceleration mechanism.
In this context, the GeV flare with a very hard spectrum in 2013
is an encouraging example as discussed in \citet{asa15}.
Future gamma-ray observations including CTA will give us
opportunity to verify the acceleration mechanism.

\acknowledgements

First we appreciate the anonymous referee for the helpful advice.
We thank Greg Madejski and Krzysztof Nalewajko for their providing SED
data points of PKS 2155-304 and PKS 1510-089.
This work is supported by Grants-in-Aid for Scientific
Research nos. 16K05291 (K.A.) and 18K03665 (M.H.) from the Ministry
of Education, Culture, Sports, Science and Technology
(MEXT) of Japan.

\end{document}